\documentclass[sigconf,authorversion,preprint]{acmart}
\AtBeginDocument{%
  \providecommand\BibTeX{{%
    \normalfont B\kern-0.5em{\scshape i\kern-0.25em b}\kern-0.8em\TeX}}}

\copyrightyear{2024}
\acmYear{2024}
\setcopyright{licensedothergov}
\acmConference[ETRA '24]{2024 Symposium on Eye Tracking Research and Applications}{June 4--7, 2024}{Glasgow, United Kingdom}
\acmBooktitle{2024 Symposium on Eye Tracking Research and Applications (ETRA '24), June 4--7, 2024, Glasgow, United Kingdom}
\acmDOI{10.1145/3649902.3653519}
\acmISBN{979-8-4007-0607-3/24/06}

\begin{document}

\title[Which Experimental Design is Better Suited for VQA Tasks?]{Which Experimental Design is Better Suited for VQA Tasks? Eye Tracking Study on Cognitive Load, Performance, and Gaze Allocations}

\author{Sita A. Vriend}
\orcid{0009-0008-8530-835X}
\affiliation{
  \institution{University of Stuttgart}
  \streetaddress{Allmandring 19}
  \country{Germany}
  \postcode{70569}
}
\email{Sita.Vriend@visus.uni-stuttgart.de}

\author{Sandeep Vidyapu}
\orcid{0000-0003-3595-5221}
\affiliation{
  \institution{University of Stuttgart}
  \streetaddress{Allmandring 19}
  \country{Germany}
  \postcode{70569}
}
\email{Sandeep.Vidyapu@visus.uni-stuttgart.de}

\author{Amer Rama}
\orcid{0000-0003-3595-5221}
\affiliation{
  \institution{University of Stuttgart}
  \streetaddress{Allmandring 19}
  \country{Germany}
  \postcode{70569}
}
\email{st156339@stud.uni-stuttgart.de}

\author{Kun-Ting Chen}
\orcid{0000-0002-3217-5724}
\affiliation{%
  \institution{University of Adelaide}
  \country{Australia}
}
\affiliation{%
  \institution{University of Stuttgart}
  \country{Germany}
}
\email{Kun-Ting.Chen@visus.uni-stuttgart.de}  

\author{Daniel Weiskopf}
\orcid{0000-0003-1174-1026}
\affiliation{%
  \institution{University of Stuttgart}
  \country{Germany}
  \postcode{70569}
}
\email{Daniel.Weiskopf@visus.uni-stuttgart.de}

\begin{abstract}
We conducted an eye-tracking user study with 13 participants to investigate the influence of stimulus-question ordering and question modality on participants using visual question-answering (VQA) tasks. We examined cognitive load, task performance, and gaze allocations across five distinct experimental designs, aiming to identify setups that minimize the cognitive burden on participants. 
The collected performance and gaze data were analyzed using quantitative and qualitative methods. Our results indicate a significant impact of stimulus-question ordering on cognitive load and task performance, as well as a noteworthy effect of question modality on task performance. These findings offer insights for the experimental design of controlled user studies in visualization research.
\end{abstract}

\ccsdesc[500]{Human-centered computing~Empirical studies in visualization}
\ccsdesc[300]{Human-centered computing~Empirical studies in HCI}

\keywords{Eye tracking, cognitive load, visual exploration, VQA, task performance, experimental design}

\begin{teaserfigure}
  \includegraphics[width=\textwidth]{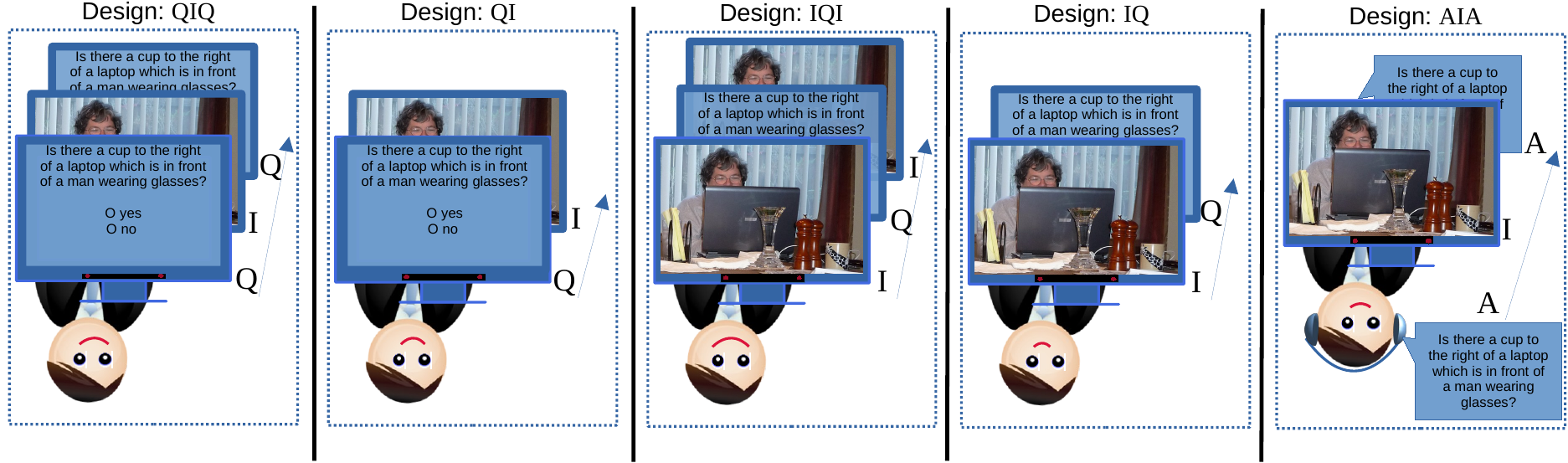}
  \caption{We investigated the effects of the order of image stimuli (I) and questions (Q), as well as the question modality (textual or auditory) for five experimental designs using visual question answering (VQA). In the QIQ, QI, IQI, and IQ designs, the question (Q) was displayed textually, whereas the participant listened to it (A) in the AIA design.} 
  \Description{Illustrative figure of five experimental design settings.}
  \label{fig:teaser}
\end{teaserfigure}

\maketitle

\section{Introduction}
Experimental designs are used to evaluate the usability and effectiveness of visualizations~\cite{purchase2012experimental}.
However, both the modality in which the tasks is presented as well as the ordering of the tasks and stimuli may influence results.

Researchers running user studies often have specific goals, such as understanding the usability of their system or visualization. They design their experiments without necessarily considering the effect task-stimulus ordering can have on users, nor might they consider the modality the task is presented in. However, these could affect cognitive load (CL), task performance, and gaze allocations. These effects might cause unwarranted results unrelated to the research aim. 

In this paper, we investigate the influence of question-stimulus order in experimental designs for visual question-answering (VQA) tasks \cite{antol2015vqa} \cite{GQA}. In addition, we examine the effects of the modality in which the question is presented (written text vs. spoken audio). For each design presented in Figure~\ref{fig:teaser}, we captured the users' gaze allocation. Subsequently, we analyzed the users' perceived CL. Finally, we performed visual exploration to understand how task-stimulus ordering and modality affect CL, performance, and gaze allocations. Our results might be useful for visualization researchers by helping them choose the best experimental design for their research goals.

\section{Related Work}\label{sec:rel}

\subsubsection*{Cognitive Load Theory (CLT)}\label{CognitiveLoadTheory}
According to CLT, the human brain has a limited capacity for information processing in the working memory~\cite{sweller1988cognitive, sweller2011cognitive} that can affect learning and performance. It is often impossible to reduce CL for all aspects of an experimental task. High CL can be inherent to a task researchers want to study. However, good experimental design, such as a suitable stimulus-task order and task modality, may reduce \textit{extraneous CL}~\cite{sweller2010element}.

\subsubsection*{Study Design}
Previous research found that factors such as stimulus characteristics, stimulus priming and information modality has an effect on task performance, CL and visual attention. Work by \citet{wang2014eye} showed that stimulus characteristics, in their case web complexity, affects attention, completion time and CL. In general, more complex websites lead to more attention, a higher CL and longer times to complete the task. Furthermore, a study by~\citet{gere2020structure} found that bottom-up factors such as stimulus size and orientation affects visual attention. A bigger size resulted in longer dwell-time and an increase in the number of fixations. Orientation influences first fixation duration and time to first fixation. 

Stimulus priming leads to more efficient gaze patterns according to \citet{castelhano2007initial}. Priming can thus facilitate visual attention guidance which, in turn, should affect task performance. 

Finally, the modality information is presented in affects visual attention. \citet{underwood2004inspecting} found that fixation duration on visual stimuli was longer than on a textual description of the visual stimuli. However, the comprehension of health information is not affected by presenting in a textual or auditory modality \cite{leroy2019comparison}. In contrast, \citet{clemens2019Comprehension} found that text comprehension and CL is affected by modality. Heads-up displays improved multitasking and reduced cognitive load compared to auditory displays. While these results demonstrate the effect of stimulus modality, we focus on the modality (textual and auditory) of the task.

According to the previous research presented, stimulus characteristics, priming and modality has an effect on user performance, visual attention and CL. However, the effect of stimulus-task order and modality in combination, while important for study designs, have not been explored. Nevertheless, we expect the results to partially carry over to our context.

\subsubsection*{Visual Search and Attention}
Visual search refers to the task of finding a target in a scene~\cite{Wolfe2021GuidedS6}. VQA provides tasks using natural language questions and natural scenes~\cite{antol2015vqa} and can be used to study visual search. 

Subsequently, \citet{AiR} analyzed gaze data to understand the impact of attention allocation, reasoning capability, and task performance. Using VQA tasks, they found that participants' visual attention was initially not accurate regarding the task but improved over time to be highly accurate.   

Several works assess CL through a combined analysis of surveys, task effectiveness, and eye-tracking data~\cite{netzel2014comparative,afzal2022investigating}. Existing eye tracking study for energy control room studies eye movement and their transitions between AOIs for assessing cognitive load~\cite{afzal2022investigating}. In this paper, we take a similar approach, focusing on a different problem of question-stimulus order's effect on participants' CL. 

\section{Research Questions}\label{sec:rq}
We investigated the influence of stimulus-task order and modality in which the task is presented using VQA tasks. We aimed to find experimental designs that reduce extraneous CL by investigating CL, performance, and gaze allocations. Accordingly, we formulated the following research questions.

\begin{description}
    \item[RQ1] Does the presentation order of image stimulus and question impact CL? Does the modality of the question affect CL?
    
    \item[RQ2] Does the presentation order of image and question impact accuracy? Does the modality of the question have an effect?

    \item[RQ3] Does the presentation order of image and question impact the gaze allocations? Does the modality of the question have an effect?
\end{description}

\section{Experimental Setup and Design}\label{sec:exp}

\begin{figure}
    \centering
    \includegraphics[width=\linewidth]{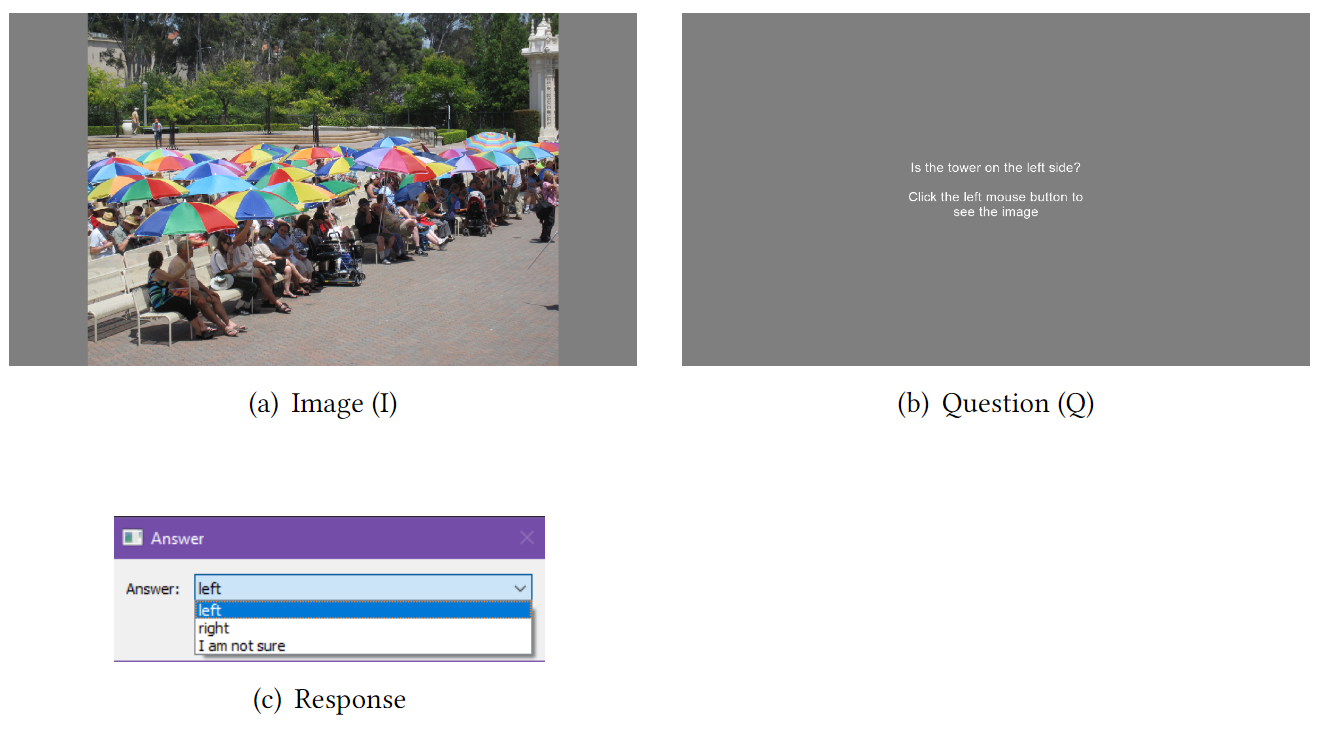}
    \caption{Example of (a) image stimulus, (b) corresponding task (question), and (c) response selection. The correct answer here is ``right'' because the tower is on the right side; however, the participant selected ``left.''}
    \label{fig:setup}
\end{figure}

\subsection{Stimuli Preparation}
We used stimuli and tasks from the GQA dataset~\cite{GQA}. To ensure the suitability of images and questions, manual quality control was performed analogously to \citet{AiR}. All images were up-scaled to a resolution of 1440\,$\times$\,1080 pixels and centered in the middle of the screen, which had a resolution of 1920\,$\times$\,1080. The background was filled with gray, as shown in Figure~\ref{fig:setup}. Text-to-speech was used for the questions in the \textit{Audio} $\longrightarrow$ Image $\longrightarrow$ Audio (AIA) design, which is defined in Subsection~\ref{sec:experimentaldesigns}. 

We chose five representative question types from the GQA dataset to represent a range of natural patterns. Example questions include: ``What|Which <type> [do you think] <is> <theObject>?'', ``Where in the scene [do you think] is <theObject> located,'' and so on. The stimuli used can be found in the supplemental material \cite{darus-3380_2023}. This includes the images for each design and training phase, as well as the related questions and answers.

\subsection{Experimental Designs and Dependent Variables}
\label{sec:experimentaldesigns}
We considered the following five experimental designs illustrated in Figure~\ref{fig:teaser}.

\begin{enumerate}
    \item[\textbf{IQ}] (Image $\longrightarrow$ Question)   
     This design is our control condition. We expected this design to lead to the highest CL and lowest accuracy since this is a free-viewing task that requires the participant to memorize the scene before knowing the question.
     
    \item[\textbf{QIQ}] (Question $\longrightarrow$ Image $\longrightarrow$ Question)
    This design shows the question twice. This might have a positive effect on CL since participants are reminded of the question.
    
    \item[\textbf{QI}] (Question $\longrightarrow$ Image)
    This is a typical visual search task. Given the question, the participant has to look for the answer in the image.
    
    \item[\textbf{IQI}] (Image $\longrightarrow$ Question $\longrightarrow$ Image)
    This design primes the participant with the image before the question. This might reduce CL because participants are already familiar with the scene when the question is posed.
    
    \item[\textbf{AIA}] (Audio $\longrightarrow$ Image $\longrightarrow$ Audio)
    This design presents the question auditorily as opposed to other designs. The participant listens to the question two times before the image is shown and one time after. Participants can dedicate attention to the image since there is no need to remember the question.    
\end{enumerate}

CL can be measured using eye tracking metrics such as fixation duration, gaze sequence, and hit-any-AOI rate (HAAR)~\cite{palinko2010estimating,MeasuringCL}. Hence, we acquired data to measure CL using multiple dependent variables: NASA Task Load Index (NASA-TLX)~\cite{NASA-TLX}, task accuracy, and gaze allocation. Participants filled out the NASA-TLX questionnaire after completing all trials of each design. Section \ref{sec:RQ1} further explains the NASA-TLX in the context of this work. In section \ref{sec:RQ2} we describe how we calculate task accuracy. Finally, we analyzed gaze allocation to understand the effect of question-stimulus order. We adopt a two-fold approach: qualitative analysis through visual scanpaths and aggregated fixation distribution, followed by both AOI- andnon-AOI-based comparative statistical analysis of gaze metrics inspired by the eye movement behavior we identified. Chen et al.~\cite{chen2023reading} used a similar analysis approach. Existing work also studies eye movement, visual scanpaths, and their transitions between AOIs for CL~\cite{afzal2022investigating}. We employed GazeAnalytics~\cite{Gazealytics2032chen} to perform both such exploratory and comparative analyses. More detailed information can be found in section \ref{sec:RQ3}.

\subsection{Apparatus and Pilot Studies}
We used a Tobii Pro Spectrum eye tracker with a sampling frequency of 600\,Hz for gaze tracking. The stimuli were displayed on a monitor of 24" diagonal size and with a screen resolution of 1920\,$\times$\,1080 and a refresh rate of 60\,Hz. A chin rest was used in front of the display monitor (approximately 63\,cm away from the screen). Participants' responses were recorded using a drop-down selection, as shown in Figure~\ref{fig:setup}(c).

Two pilot studies were conducted before the actual study. The first pilot study examined the procedure and implementation of the study and served to identify ambiguous instructions. The second pilot study aimed to estimate the study duration.

The first pilot study was conducted with two participants who had prior knowledge of the study. A shorter version of the experiment was used in which each design block consisted of three question-image pairs. We identified and corrected unclarities in the pre-test survey and the experiment instructions. 

Using the intended study setup, we conducted a second pilot study with one participant. The time measured for this pilot study included time for the preparations, such as the participant reading through the study description and signing the consent form. The measured time was about 35 minutes. We used this as an estimated time for the study description and in the invitations for the study.

\subsection{Participants and Experimental Procedure}
The 13 participants (10 male, 2 female, 1 other) volunteered. Their age ranged from 18 to 29 years. All participants had normal or corrected vision and dominant ``left to right'' reading behavior. Two participants had red-green color blindness. 

We performed a counter-balanced within-subject experimental procedure \cite{brooks2012counterbalancing}. Initially, a demographics survey was conducted, followed by a briefing. A training session followed with three image-question pairs, upon which eye tracking calibration and validation were done. Lastly, the trials of 20 question-image pairs per design were conducted. Each participant filled in the NASA-TLX questionnaire after each block with a design.

\section{Results and Analysis}\label{sec:res}
We excluded one participant due to incomplete tasks for the AIA design. In total, we analyzed the data from 12 participants. The data and R scripts used for the analysis can be found in the supplemental material \cite{darus-3380_2023}.

\subsection{RQ1: Does the presentation order of image stimulus and question impact CL? Does the modality of the question affect CL?}
\label{sec:RQ1}

\begin{figure}
    \centering
    \includegraphics[width=\linewidth]{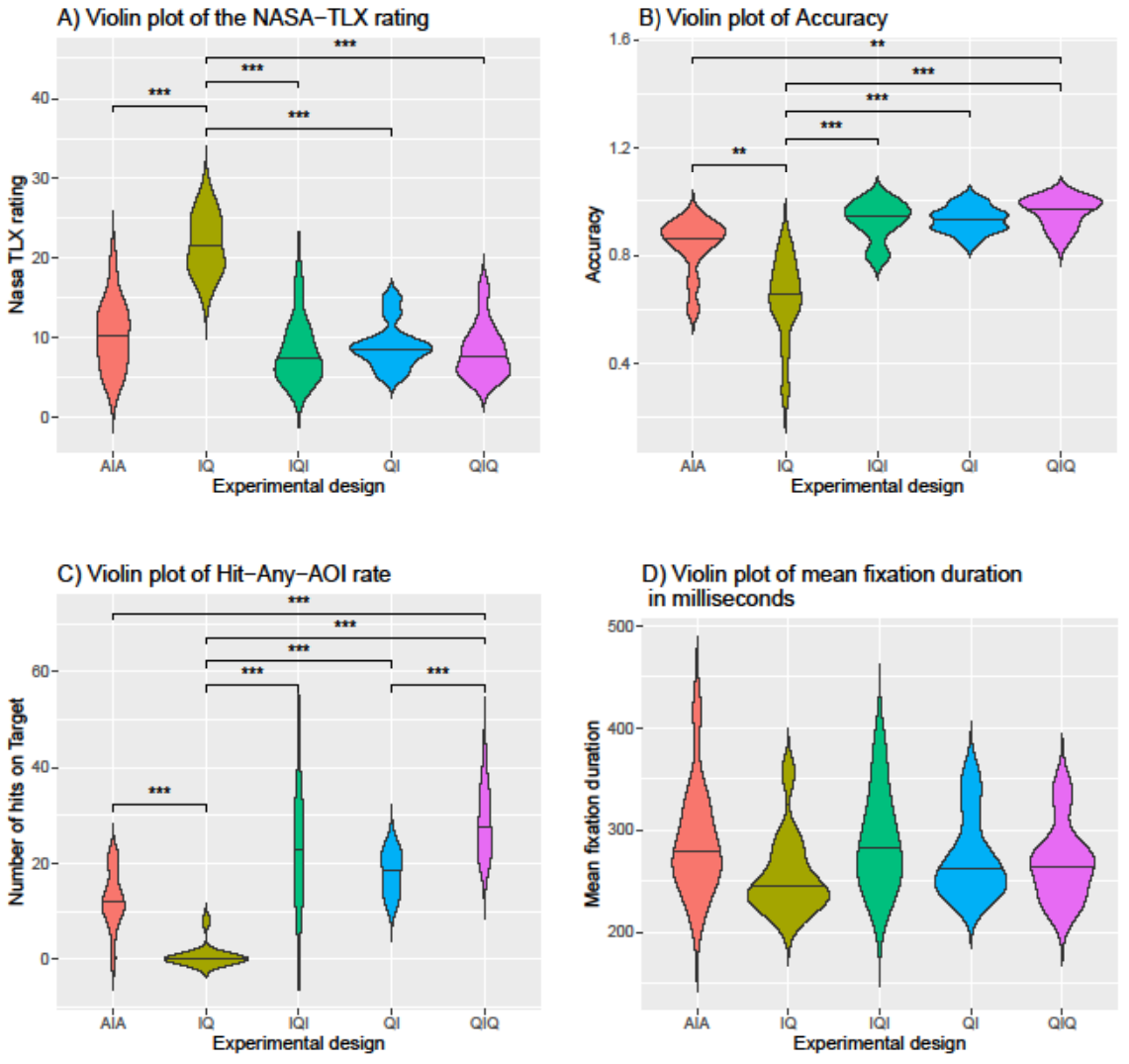}
    \caption{Violin plots of cognitive load according to NASA-TLX rating (A), task accuracy (B), hit-any-AOI rate per experimental design (C), and mean fixation duration measured in milliseconds (D). The horizontal black line in each plot represents the mean. Significant differences according to post-hoc tests are marked with asterisks (* \emph{p} < 0.05; ** \emph{p} < 0.01; *** \emph{p} < 0.001).}
    \label{fig:descriptive_all}
\end{figure}

We evaluated the subjective experience of CL using the NASA-TLX Index. The scores were calculated by adding up the responses of the following sub-scales per design and participant: mental demand, temporal demand, effort, frustration, and performance. The ratings were not normally distributed according to the Shapiro-Wilk test $(W = 0.87, p < 0.001)$. Hence, the Friedman test was used to examine the data. This test indicated statistically significant differences $(\chi^2(4) = 27.095, p < 0.001)$ with a large effect size $(w = 0.616)$. 
The Bonferroni-corrected post-hoc test results in \autoref{fig:descriptive_all} show that the IQ design had the highest subjective experience of CL. There was no statistically significant difference between the other designs, including the AIA.

\subsection{RQ2: Does the presentation order of image and question impact accuracy? Does the modality of the question have an effect?}
\label{sec:RQ2}
Task accuracy was calculated by dividing the number of correct answers by the total number of questions in a design. Since the data of task accuracy was not normally distributed $(W = 0.82, p = 0.006)$, a Friedman test was employed. Accuracy between different designs was found to be statistically significant $(\chi^2(4) = 32, p < 0.001)$, with a large effect size $(w = 0.727)$. 

\autoref{fig:descriptive_all} shows violin plots with the results of the Bonferroni-adjusted post-hoc tests. The IQ performed significantly worse compared to other designs. We also found a statistically significant difference between AIA, the design with the second lowest accuracy, and QIQ, which had the highest accuracy.

\subsection{RQ3: Does the presentation order of image and question impact the gaze allocations? Does the modality of the question have an effect?} 
\label{sec:RQ3}

To answer this question, we analyzed gaze data on the images only. In regards to the IQI design, we only analyze the gaze data of the second image shown.

The fixation detection parameters are based on a dispersion-based I-DT algorithm by \citet{salvucci2000identifying}. After a preliminary analysis of plotting raw gaze points for each stimulus, we opt for a maximum dispersion of 130 pixels, defined by a bounding box $([\max(x) - \min(x)] + [\max(y) - \min(y)])$ for a cluster of gaze points and a minimum fixation clustering time window of 80\,ms. This resulted in 10--20 fixations per participant and stimulus. A similar configuration was used in predicting VQA human scanpaths~\cite{chen2021predicting}.

\subsubsection{Scanpaths and Aggregated Fixation Distribution}
\label{sec:scanpaths}
\begin{figure}
    \centering     
    \includegraphics[angle=90, width=\linewidth]{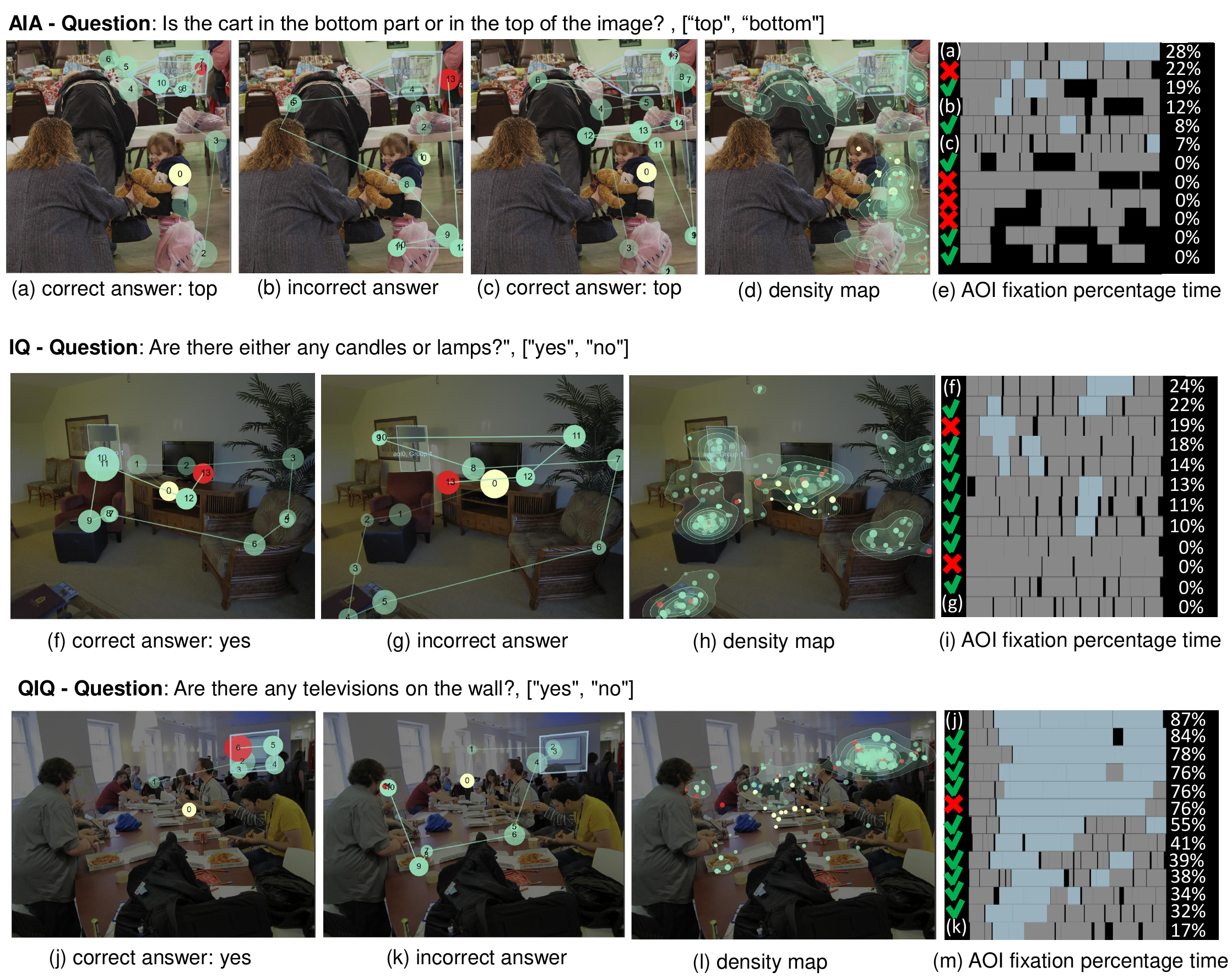}    
    \Description[]{}
    \caption[]{Visual scanpath overlaid on images of three study designs of a selected question, where the number and radius indicate the fixation sequence and its duration, respectively. 
    The yellow and red dots indicate the beginning and the end of a scanpath. 
    Aggregated attention is displayed in density maps (d, h, l), while scarf plots (e, i, m) show the fixation duration over a target AOI (colored in light blue) across all participants. 
    The task correctness is shown as green ticks (for correct answers) and red crosses (for incorrect answers). 
    Percentage shows the relative fixation duration spent on the target AOI.
    Scarf plots in each row are ordered by decreasing relative fixation duration of a target AOI.
    \label{fig:scanpath_strategies}
    }
\end{figure}

\autoref{fig:scanpath_strategies} shows the results of the analysis of aggregated fixation distribution and scanpaths of a representative instance for each question-stimuli order. The gaze allocation plot results are shown for the image reading phase (I) for AIA, IQ, and QIQ. For AIA, we found limited gaze allocation to the target (colored in light blue in~\autoref{fig:scanpath_strategies}(e)) asked by the question. \autoref{fig:scanpath_strategies}(a) shows that the target (cart) was fixated for a short duration. It was rarely fixated by about half of the participants, as seen in \autoref{fig:scanpath_strategies}(e). The scanpath results also show that there is a lower number of fixations over the target for AIA, compared with QIQ. We found similar behavior in other AIA stimuli, where participants did not pay attention to the question's target. \autoref{fig:scanpath_strategies}(d) shows that the aggregated attention map shows denser fixations over the right of the child, and to the right of the cart, but not obviously over the target. This could explain its lower task accuracy results compared to QIQ, QI, and IQI.

We also note that there were more gaps (blank areas in \autoref{fig:scanpath_strategies}(e)) between fixations where no gaze points were detected by the eye tracker. Participants may gazed out of the screen at times for an auditory stimulus. 

For IQ, aggregated attention maps show at least five distinct clusters of fixations (\autoref{fig:scanpath_strategies}(h)). This study design involves visual search based on memory. Participants may needed to quickly gain an overview of the scene, before they are asked the question.

For QIQ, IQ, and IQI, we observed similar gaze allocations under the same question type. \autoref{fig:scanpath_strategies}(l) shows an apparent cluster of fixations relevant to the areas required to answer the question. The relative duration scarf plot of fixations over the target shows a range of low to high percentages of fixation duration over targets, indicating that gaze allocation varies by reading strategies. \autoref{fig:scanpath_strategies}(j) shows a correct answer using a high concentration of fixations on the wall after the beginning of scanpaths. Also for the correct answer, other participants used a different strategy, spending 32\% to 55\% time inspecting the wall, while also looking around, and back to visit the wall again (see scarf plots). There was also a clear depth of fixations over the target (longer fixation duration over the target).

\subsubsection{Comparative Statistical Analysis Based on Gaze Metrics}
Inspired by scanpaths results and aggregated fixation distribution (\autoref{sec:scanpaths}), we quantify our insights by defining AOIs by the target mentioned in the question. For the images, the hit-any-AOI Rate (HAAR) was calculated by the number of fixations over AOIs divided by the total number of fixations per image per participant~\cite{wang2022impact}. HAAR can be used to compare the relative gaze allocation over AOIs across participants. 
Most stimuli had a target mentioned by the question, thus one AOI could be defined, except for questions where no target is available, such as 'Does the image contain a chair (where the image does not have a chair). In such cases, the gaze allocation results for these stimuli was excluded from the analysis. Example AOIs are shown in \autoref{fig:scanpath_strategies}.

We aggregated HAAR by calculating the median HAAR for each combination of participant and experimental design. To perform a statistical significant test, we ensured the number of samples was equal by randomly dropping some samples. Since the data did not meet the assumptions for ANOVA for the IQ design $(W = 0.33, p < 0.001)$, a Friedman rank sum test was done. The results show a significant difference in HAAR between the designs $(\chi^2(4) = 36.69, p < 0.001)$ with a large effect size $(w = 0.77)$. A Bonferroni-corrected post-hoc test was done and visualized in \autoref{fig:descriptive_all}. As shown, IQ's HAAR was lowest. When a participant's gaze hit the target in IQ, it was likely at random. The AIA design had the second lowest HAAR but was only significantly lower than QIQ. We confirmed our finding from the exploratory analysis of Section~\ref{sec:RQ3}. QIQ's HAAR was significantly higher compared to all other designs, except IQI. 

Furthermore, similar to Netzel et al.~\cite{netzel2014comparative}, we investigated mean fixation duration to assess CL. Mean fixation duration independent of AOI was analyzed to complement our findings from the NASA-TLX ratings (i.e., subjective experience of CL) because fixation duration can be an indirect measure of CL~\cite{chen2011eye}. 
The mean fixation duration was aggregated by calculating the mean of the mean fixation duration for each combination of participant and experimental design. The data was not normally distributed $(W = 0.92, p < 0.001)$, hence we applied a log transformation, which made the data approximately normally distributed $(W = 0.96, p = 0.02)$. A one-way ANOVA showed a significant difference in log-transformed mean fixation duration between the designs after sphericity corrections $(F(2.52, 27.74)=4.619, p[GG]=0.013)$, with a medium effect size $(\eta^2[g] = 0.088)$. The results of the Bonferroni-corrected post-hoc tests can be found in \autoref{fig:descriptive_all}). Even though the IQ design had the lowest mean fixation duration, there were no significant differences with the other designs.

\section{Discussion and Conclusion}\label{sec:disc}
This work explored five experimental designs for user studies, intending to optimize experimental design. Our approach was to alter the image-question order and the modality of the question. We collected NASA-TLX ratings, task accuracy, and gaze data per design in a user study. The results suggest some significant differences between different presentation orders. 

In general, we found that all experimental designs perform better than the control, IQ design. IQ was mentally taxing since it asked a lot from participants´ memory. This means that it is advisable to present the question at least once before the image.
While not statistically significant, the auditory modality seems to lead to higher CL and lower accuracy. The presentation order of AIA and QIQ are the same, so we expected performance to be similar. However, the question in QIQ is written rather than spoken. An explanation could be that the text allowed participants to read and comprehend at their own pace. People can read and re-read quickly even though the question was only briefly shown. On the other hand, people cannot re-listen at their speed, which could lead some to struggle to comprehend the task. However, it is worth noting that AIA was the only design that presented the question in an auditory fashion. 

QI might be the best design. The accuracy and CL are similar between QI, IQI, and QIQ. The gaze data showed some interesting differences. While the IQ leads to one of the smallest variances in HAAR, the IQI leads to the largest variance (see \autoref{fig:descriptive_all}C). Such a large variance indicates inconsistency; some participants' gaze did not hit the target, while others hit the target very often. The smaller variance in IQ indicates a more consistent gaze performance. Furthermore, IQI shows the image twice and QIQ presents the question twice. This takes three seconds longer per task compared to QI. Hence QI is more efficient as well.

\subsubsection*{Limitations}
The images and tasks in the VQA set are natural and thus varied. There were tasks about a target's characteristics, the presence in the scene, and location. Saliency aspects of targets such as size and location also varied. Task type and scene guidance by context influence search strategies \cite{Wolfe2021GuidedS6}. While task types were distributed relatively evenly between the conditions, images were not.
%
Due to the nature of the stimuli, we were not able to reliably measure pupil dilation as a measure of CL. Hence, we used gaze allocation metrics which are known to be indirect measures of CL: Hit-any-AOI-rate and mean fixation duration. We additionally employed the widely used NASA-TLX, a standardized questionnaire known to reliably measure the subjective experience of CL.
%
Our study tested the designs within-subject which could have led to a learning effect for the textual modality. Participants got more exposure to the textual modality since AIA was the only auditory modality. This might have led to poor performance in the AIA design.

\subsubsection*{Future work}
Our recommendations for study designs are based on the VQA database, which contains image-question pairs of natural scenes. Future research could extend our work to different visual tasks. For example, less natural, more controlled stimuli could be used as a basis for visualization research. 
Additionally, future research could explore other experimental designs. It would be interesting to see research further investigating question modality. A direct auditory competitor to QI, AI could be explored for example.

\begin{acks}
We would like to thank the participants for their time and effort. This work was funded by the Deutsche Forschungsgemeinschaft (DFG, German Research Foundation)---Project ID 251654672---TRR 161 (Project A01)
\end{acks}

\bibliographystyle{ACM-Reference-Format}
\bibliography{VQA_lit}

\end{document}